\documentclass[12pt]{article}
\usepackage{graphicx}
\usepackage{lscape}
\usepackage{citesort}
\usepackage{amssymb}
\usepackage{appendix}
\usepackage{multirow}

\begin{document}
\def\qq{\langle \bar q q \rangle}
\def\uu{\langle \bar u u \rangle}
\def\dd{\langle \bar d d \rangle}
\def\sp{\langle \bar s s \rangle}
\def\GG{\langle g_s^2 G^2 \rangle}
\def\Tr{\mbox{Tr}}
\def\figt#1#2#3{
        \begin{figure}
        $\left. \right.$
        \vspace*{-2cm}
        \begin{center}
        \includegraphics[width=10cm]{#1}
        \end{center}
        \vspace*{-0.2cm}
        \caption{#3}
        \label{#2}
        \end{figure}
	}
	
\def\figb#1#2#3{
        \begin{figure}
        $\left. \right.$
        \vspace*{-1cm}
        \begin{center}
        \includegraphics[width=10cm]{#1}
        \end{center}
        \vspace*{-0.2cm}
        \caption{#3}
        \label{#2}
        \end{figure}
                }

\def\ds{\displaystyle}
\def\beq{\begin{equation}}
\def\eeq{\end{equation}}
\def\bea{\begin{eqnarray}}
\def\eea{\end{eqnarray}}
\def\beeq{\begin{eqnarray}}
\def\eeeq{\end{eqnarray}}
\def\ve{\vert}
\def\vel{\left|}
\def\ver{\right|}
\def\nnb{\nonumber}
\def\ga{\left(}
\def\dr{\right)}
\def\aga{\left\{}
\def\adr{\right\}}
\def\lla{\left<}
\def\rra{\right>}
\def\rar{\rightarrow}
\def\lrar{\leftrightarrow}  
\def\nnb{\nonumber}
\def\la{\langle}
\def\ra{\rangle}
\def\ba{\begin{array}}
\def\ea{\end{array}}
\def\tr{\mbox{Tr}}
\def\ssp{{\Sigma^{*+}}}
\def\sso{{\Sigma^{*0}}}
\def\ssm{{\Sigma^{*-}}}
\def\xis0{{\Xi^{*0}}}
\def\xism{{\Xi^{*-}}}
\def\qs{\la \bar s s \ra}
\def\qu{\la \bar u u \ra}
\def\qd{\la \bar d d \ra}
\def\qq{\la \bar q q \ra}
\def\gGgG{\la g^2 G^2 \ra}
\def\q{\gamma_5 \not\!q}
\def\x{\gamma_5 \not\!x}
\def\g5{\gamma_5}
\def\sb{S_Q^{cf}}
\def\sd{S_d^{be}}
\def\su{S_u^{ad}}
\def\sbp{{S}_Q^{'cf}}
\def\sdp{{S}_d^{'be}}
\def\sup{{S}_u^{'ad}}
\def\ssp{{S}_s^{'??}}

\def\sig{\sigma_{\mu \nu} \gamma_5 p^\mu q^\nu}
\def\fo{f_0(\frac{s_0}{M^2})}
\def\ffi{f_1(\frac{s_0}{M^2})}
\def\fii{f_2(\frac{s_0}{M^2})}
\def\O{{\cal O}}
\def\sl{{\Sigma^0 \Lambda}}
\def\es{\!\!\! &=& \!\!\!}
\def\ap{\!\!\! &\approx& \!\!\!}
\def\ar{&+& \!\!\!}
\def\ek{&-& \!\!\!}
\def\kek{\!\!\!&-& \!\!\!}
\def\cp{&\times& \!\!\!}
\def\se{\!\!\! &\simeq& \!\!\!}
\def\eqv{&\equiv& \!\!\!}
\def\kpm{&\pm& \!\!\!}
\def\kmp{&\mp& \!\!\!}
\def\mcdot{\!\cdot\!}
\def\erar{&\rightarrow&}

% .........................................................

\def\simlt{\stackrel{<}{{}_\sim}}
\def\simgt{\stackrel{>}{{}_\sim}}

% .........................................................

\renewcommand{\textfraction}{0.2}    %float (figures) parameters
\renewcommand{\topfraction}{0.8}   

\renewcommand{\bottomfraction}{0.4}   
\renewcommand{\floatpagefraction}{0.8}
\newcommand\mysection{\setcounter{equation}{0}\section}

\def\baeq{\begin{appeq}}     \def\eaeq{\end{appeq}}  
\def\baeeq{\begin{appeeq}}   \def\eaeeq{\end{appeeq}}
\newenvironment{appeq}{\beq}{\eeq}   
\newenvironment{appeeq}{\beeq}{\eeeq}
\def\bAPP#1#2{
 \markright{APPENDIX #1}
 \addcontentsline{toc}{section}{Appendix #1: #2}
 \medskip
 \medskip
 \begin{center}      {\bf\LARGE Appendix #1 :}{\quad\Large\bf #2}
% \begin{center}      {\bf\LARGE Appendix  :}{\quad\Large\bf #2}
\end{center}
 \renewcommand{\thesection}{#1.\arabic{section}}
\setcounter{equation}{0}
        \renewcommand{\thehran}{#1.\arabic{hran}}
\renewenvironment{appeq}
  {  \renewcommand{\theequation}{#1.\arabic{equation}}
     \beq
  }{\eeq}
\renewenvironment{appeeq}
  {  \renewcommand{\theequation}{#1.\arabic{equation}}
     \beeq
  }{\eeeq}
\nopagebreak \noindent}

\def\eAPP{\renewcommand{\thehran}{\thesection.\arabic{hran}}}

\renewcommand{\theequation}{\arabic{equation}}
\newcounter{hran}
\renewcommand{\thehran}{\thesection.\arabic{hran}}

\def\bmini{\setcounter{hran}{\value{equation}}
\refstepcounter{hran}\setcounter{equation}{0}
\renewcommand{\theequation}{\thehran\alph{equation}}\begin{eqnarray}}
\def\bminiG#1{\setcounter{hran}{\value{equation}}
\refstepcounter{hran}\setcounter{equation}{-1}
\renewcommand{\theequation}{\thehran\alph{equation}}
\refstepcounter{equation}\label{#1}\begin{eqnarray}}

%       the stuff below defines \eqalign and \eqalignno in such a
%       way that they will run on Latex

\newskip\humongous \humongous=0pt plus 1000pt minus 1000pt
\def\caja{\mathsurround=0pt}
%\def\eqalign#1{\,\vcenter{\openup1\jot
%\caja   %\ialign{\strut \hfil$\displaystyle{##}$&$
%\displaystyle{{}##}$\hfil\crcr#1\crcr}
%}\,}

% ...........................................................

\title{
         {\Large
                 {\bf
Heavy baryon--light vector meson couplings in QCD 
                 }
         }
      }

\author{\vspace{1cm}\\
{\small T. M. Aliev \thanks {e-mail:
taliev@metu.edu.tr}~\footnote{permanent address:Institute of
Physics,Baku,Azerbaijan}\,\,, K. Azizi \thanks {e-mail:
kazizi@dogus.edu.tr}\,\,, M. Savc{\i} \thanks
{e-mail: savci@metu.edu.tr}} \\
{\small Physics Department, Middle East Technical University,
06531 Ankara, Turkey }\\
{\small$^\ddag$ Physics Division,  Faculty of Arts and Sciences,
Do\u gu\c s University,} \\
{\small Ac{\i}badem-Kad{\i}k\"oy,  34722 Istanbul, Turkey}}

\date{}

\begin{titlepage}
\maketitle
\thispagestyle{empty}

\begin{abstract}
The strong coupling constants of heavy baryons with light vector mesons are
calculated in the framework of the light cone QCD sum rules using the most
general form of the interpolating currents for the heavy baryons. It is shown
that the sextet--sextet, sextet--antitriplet and antitriplet--antitriplet
transitions are described by one invariant function for each class of
transitions. The values of the electric and magnetic coupling constants 
for these transitions are obtained.
\end{abstract}

%\vspace{1cm}
~~~PACS number(s): 11.55.Hx, 13.75.Gx, 13.75.Jz
\end{titlepage}

\section{Introduction}

Experimental and theoretical studies of charmed and bottom baryons are
recognized to be one of the main research areas in particle physics. 
During the last few years quite interesting experimental observations 
have been obtained in heavy hadron spectroscopy. With the help of several 
refined measurements, the new states are discovered both in the charm and 
bottom sector (for recent experimental results see review \cite{Rhv01}). 
These observations come from both the BaBar and BELLE, as
well as from CDF and D$\rlap/$O Collaborations. LHC opens the possibilities
for the discovery and detailed study of the new baryon states \cite{Rhv02}.
The considerable progress on the experimental side, has stimulated the
theoretical investigation for understanding the dynamics of heavy flavor
hadrons.
Careful and comprehensive theoretical studies of the experimental results on
heavy hadron spectroscopy and analysis of their weak and strong decays can provide essential
knowledge on the quark structure of these hadrons. The strong coupling
constants of the light pseudoscalar and vector mesons with heavy baryons are
the main parameters for understanding the dynamics of heavy baryons. For this
reason, reliable determination of the strong coupling constants of light
pseudoscalar and vector mesons with heavy baryons within QCD receives
special attention. Unfortunately, at the hadronic scale, QCD becomes
nonperturbative and it makes impossible to calculate these coupling
constants starting from QCD Lagrangian. Therefore, for calculation of these
couping constants some nonperturbative methods are needed. The QCD sum rule
approach, which is based on the fundamental QCD Lagrangian, is one of the
most attractive and applicable approaches \cite{Rhv03}. In this work, we calculate the strong
coupling constants of light vector mesons with heavy baryons within the
light cone version of the QCD sum rules (LCSR) (for more about LCSR see
\cite{Rhv04}). The coupling constants of pseudoscalar mesons with heavy
baryons is studied in detail in the same framework in \cite{Rhv05}.

The work is arranged in the following way. In section 2, the light cone sum
rules for the coupling constants of light vector mesons with heavy baryons
are obtained. In the following section, the numerical analysis of the
obtained sum rules is performed and a comparison of our results with ones
existing in literature is presented.

\section{Light cone QCD sum rules for the heavy baryon-- light vector meson
couplings} 

In this section, we calculate the strong coupling constants of light--vector
mesons with heavy baryons. Before starting to calculate these coupling
constants few words about the $SU(3)_f$ classification of the heavy baryons
are in order. Baryons with a heavy single quark belong to either
antisymmetric antitriplet $\bar{3}_F$ or symmetric sextet $6_F$
representations. Using the symmetry properties of the wave function, the spin of
the light diquark is equal to zero for the antitriplet, while it is equal to
one for the sextet. For this reason the total spin of the ground state
baryons is $1/2$ for $\bar{3}_F$, but it can be both $3/2$ and $1/2$ for
$6_F$. In the present work we restrict ourselves by considering only spin
$1/2$ heavy baryons.

In order to obtain the strong coupling constants of light--vector mesons
with heavy baryons within the LCSR method, we consider the following
correlation function:
\bea
\label{ehv01}
\Pi^{(ij)} = i \int d^4x e^{ipx} \lla V(q) \vel {\cal T} \left\{
\eta^{(i)}_{B_2} (x) \bar{\eta}^{(j)}_{B_1} (0) \right\} \ver 0 \rra~,
\eea
where the indices $i$ and $j$ get two values and describe the sextet--sextet 
($i=1,~j=1$), sextet--triplet ($i=1,~j=2$) and triplet--triplet ($i=2,~j=2$) 
transitions, respectively. In further respect, we introduce the definitions,
$\Pi^{(11)} = \Pi^{(1)}$, $\Pi^{(12)} = \Pi^{(2)}$ and $\Pi^{(22)} =
\Pi^{(3)}$, correspondingly. Note that, the $V(q)$ in Eq. (\ref{ehv01})
denotes the light--vector mesons $(\rho,~\omega,~K^\ast,~\phi)$ with momentum
$q$ and $\eta$ is the interpolating current of the heavy baryon.

The correlation function (\ref{ehv01}) can be calculated in two different
kinematical regions, namely, in terms of hadrons (phenomenological side), as
well as in deep euclidean region when $p^2 \rar - \infty$, in quark and
gluon degrees of freedom using operator product expansion (OPE) (theoretical
side). Equating both representations with the help of dispersion relations
allows us to obtain the sum rules that will be used for calculation of the
strong coupling constants of the light--vector mesons with heavy baryons.

The phenomenological part of the correlation function can be obtained by
saturating it with hadrons which carry the same quantum numbers as their interpolating currents. Separating contribution of the ground state
baryons we get,
\bea
\label{ehv02}
\Pi^{(ij)} =  {\lla 0 \vel \eta^{(i)}_{B_2} \ver
B_2(p) \rra \lla B_2(p) V(q) \vel \right.
B_1(p+q) \rra \lla B_1(p+q) \vel \bar{\eta}^{(j)} _{B_1}\ver
0 \rra \over \ga p^2-m_2^2 \dr \left[(p+q)^2-m_1^2\right]}
+ \cdots~,
\eea
where $m_1$ and $m_2$ are the masses of the initial and final baryons and
dots represent contributions from higher states and continuum.
The matrix elements appearing in Eq. (\ref{ehv02}) are defined as follows:
\bea
\label{ehv03}
\lla 0 \vel \eta^{(i)}_{B} \ver B(p) \rra \es \lambda_i \bar{u}(p)~, \\
\label{ehv04}
\lla B_2(p) V(q) \vel \right. B_1(p+q) \rra \es \bar{u}(p) \Bigg[
f_1 \gamma_\mu - i f_2 \sigma_{\mu\nu} q^\nu {1\over m_1+m_2} \bigg] u(p+q)
\varepsilon^\mu~,
\eea
where $\lambda_i$ are the residues of the heavy baryons and $\varepsilon^\mu$ are the momentum and vector polarization of the
vector meson, $f_1$ and $f_2$ are the charge and magnetic form factors,
respectively.

Performing summation over spins of the baryons and using Eqs.
(\ref{ehv02})--(\ref{ehv04}), we get for the phenomenological part,
\bea
\label{ehv05}
\Pi^{(ij)} \es i {\lambda_i \lambda_j \over \ga p^2-m_2^2 \dr
\left[(p+q)^2-m_1^2\right]} \Big\{ (f_1+f_2) 
\rlap/p \rlap/\varepsilon \rlap/q + 2 f_1 
(\varepsilon\!\cdot\! p) \rlap/p
+ \mbox{\rm other structures} \Big\}~.
\eea
The reason why we choose the structures $\rlap/p \rlap/\varepsilon \rlap/q $ and
$(\varepsilon\!\cdot\! p) \rlap/p$ is that they show the best convergence. 

In calculating the theoretical part from the QCD side, explicit expressions
of the interpolating currents of heavy baryons are needed. Using the fact
that the sextet (antitriplet) current should be symmetric (antisymmetric)
with respect to the light quarks, the most general form of the interpolating
currents for the spin--$1/2$ sextet and antitriplet baryons can be written
in the following form,
\bea
\label{ehv06}
\eta_Q^{(s)} \es - {1\over \sqrt{2}} \epsilon^{abc} \Big\{ \Big( q_1^{aT}
C Q^b \Big) \gamma_5 q_2^c - \Big( Q^{aT} C q_2^b \Big) \gamma_5 q_1^c +
\beta \Big[ \Big( q_1^{aT} C \gamma_5 Q^b \Big) q_2^c - \Big( Q^{aT} C
\gamma_5 q_2^b \Big) q_1^c \Big] \Big\}~, \nnb \\
\eta_Q^{(a)} \es {1\over \sqrt{6}} \epsilon^{abc} \Big\{ 2 \Big( q_1^{aT}
C q_2^b \Big) \gamma_5 Q^c + \Big( q_1^{aT} C Q^b \Big) \gamma_5 q_2^c +
\Big( Q^{aT} C q_2^b \Big) \gamma_5 q_1^c \nnb \\
\ar \beta \Big[2 \Big( q_1^{aT} C \gamma_5 q_2^b \Big) Q^c
+ \Big(q_1^{aT} C \gamma_5 Q^b \Big) q_2^c +
\Big(Q^{aT} C \gamma_5 q_2^b \Big) q_1^c \Big]\Big\}~,
\eea
where the superscripts $s$ and $a$ refer to the sextet and antitriplet,
respectively, and subscripts $a,b,c$ are the color indices, $C$ is the
charge conjugation operator, $\beta$ is an arbitrary parameter, and
$\beta=-1$ case describes the Ioffe current. The light quark fields 
$q_1$ and $q_2$ for the sextet and antitriplet are given in Table 1.

% .........................................................

\begin{table}[h]

\renewcommand{\arraystretch}{1.3}
\addtolength{\arraycolsep}{-0.5pt}
\small
$$
\begin{array}{|l|c|c|}
\hline \hline
                       & q_1 & q_2 \\  \hline
 \Sigma_{b(c)}^{+(++)} & u   & u  \\
 \Sigma_{b(c)}^{0(+)}  & u   & d  \\
 \Sigma_{b(c)}^{-(0)}  & d   & d  \\
 \Xi_{b(c)}^{-(0)'}    & d   & s  \\
 \Xi_{b(c)}^{0(+)'}    & u   & s  \\
 \Omega_{b(c)}^{-(0)}  & s   & s  \\
 \Lambda_{b(c)}^{0(+)} & u   & d  \\
 \Xi_{b(c)}^{-(0)}     & d   & s  \\
 \Xi_{b(c)}^{0(+)}     & u   & s  \\
\hline \hline
\end{array}
$$
\caption{The light quarks $q_1$ and $q_2$ for the sextet and
antitriplet baryons}
\renewcommand{\arraystretch}{1}
\addtolength{\arraycolsep}{-1.0pt}
\end{table}

Before calculating the correlation functions responsible for  all sextet--sextet--vector mesons
(SSV), sextet--antitriplet--vector mesons (SAV) and antitriplet--antitriplet--vector mesons (AAV) transitions
 from the QCD side, we 
  find the
relations among invariant functions for each of the above--mentioned
classes (see also \cite{Rhv05,Rhv06,Rhv07,Rhv08,Rhv09,Rhv10}). As a result, we find that the magnetic and electric couplings
for each class of transitions are described in terms of only one invariant
function. Of course, the invariant functions for each class of transitions
are different from each other in the general case. It should also be noted
that the relations among the invariant functions are all structure
independent.

After these remarks, we can proceed to establish relations among
invariant functions involving the couplings of  sextet--sextet transitions.
Let us first consider $\Sigma_b^0 \rar \Sigma_b^0 \rho^0$ transition. The
invariant function responsible for this transition can schematically be
written as:
\bea
\label{ehv07}
\Pi^{\Sigma_b^0 \rar \Sigma_b^0 \rho^0} = g_{\rho\bar{u}u} \Pi_1^{(1)}(u,d,b) +
g_{\rho\bar{d}d} \Pi_1^{'(1)}(u,d,b) + g_{\rho\bar{b}b} \Pi_2^{(1)}(u,d,b)~,
\eea   
where we have introduced the formal notations,
\bea
\label{ehv08}
\Pi_1^{(1)}(u,d,b) \es \lla \bar{u}u \vel \Sigma_b^0 \bar{\Sigma}_b^0 \ver 0 \rra~, \nnb \\
\Pi_2^{(1)}(u,d,b) \es \lla \bar{b}b \vel \Sigma_b^0 \bar{\Sigma}_b^0 \ver 0 \rra~, \nnb \\
\Pi_1^{'(1)}(u,d,b) \es \lla \bar{d}d \vel \Sigma_b^0 \bar{\Sigma}_b^0 \ver 0 \rra~.
\eea
The interpolating current of $\rho^0$ is formally written in the form,
\bea
\label{ehv09}
J^{\rho^0} _\mu= \sum_{u,d,b} g_{\rho\bar{q}q} \bar{q} \gamma_\mu q~,
\eea
where we have set $ g_{\rho^0\bar{b}b}=0$  and $ g_{\rho^0\bar{u}u}= -
g_{\rho^0\bar{d}d}=1/\sqrt{2}$. Physically, each term on the
right hand side of Eq. ({\ref{ehv07}) describes emission of the $\rho^0$ meson
from $u$, $d$ and $b$ quarks of the $\Sigma_b^0$ baryon, respectively. Since
interpolating current of $\Sigma_b^0$ is symmetric under the exchange of $u$
and $d$ quarks, it leads to the result, $\Pi_1^{'(1)}(u,d,b)=\Pi_1^{(1)}(d,u,b)$.
Using the definitions given in Eq. (\ref{ehv08}) and taking the remark after
Eq. (\ref{ehv09}) into consideration, we obtain,
\bea
\label{ehv10}     
\Pi^{\Sigma_b^0 \rar \Sigma_b^0 \rho^0} = {1\over \sqrt{2}}
\Big[\Pi_1^{(1)}(u,d,b) - \Pi_1^{(1)}(d,u,b) \Big]~.
\eea
Obviously, in the $SU(2)_f$ limit $\Pi^{\Sigma_b^0 \rar \Sigma_b^0 \rho^0}
=0$. 

The invariant function describing $\Sigma_b^+ \rar \Sigma_b^+ \rho^0$
($\Sigma_b^- \rar \Sigma_b^- \rho^0$) can be obtained from the invariant
function for the $\Sigma_b^0 \rar \Sigma_b^0 \rho^0$ transition with the
help of the replacement, $d \rar u$ ($u \rar d$), and using the fact that
$\Sigma_b^0 = -\sqrt{2} \Sigma_b^+$ ($\sqrt{2} \Sigma_b^-$). As a result, we
obtain,
\bea
\label{ehv11}
4 \Pi_1^{(1)}(u,u,b) \es -2 \lla \bar{u} u \vel \Sigma_b^+ \bar{\Sigma}_b^+
\ver 0 \rra~, \\
\label{ehv12}
4 \Pi_1^{(1)}(d,d,b) \es 2 \lla \bar{d} u \vel \Sigma_b^- \bar{\Sigma}_b^- 
\ver 0 \rra~.
\eea
Since $\Sigma_b^{+(-)}$ contains two $u(d)$ quarks, there are four possible
ways for emitting $\rho^0$ from the $u(d)$ quark. Hence, the related invarian functions are obtained as:
\bea
\label{ehv11111}
\Pi^{\Sigma_b^+ \rar \Sigma_b^+ \rho^0} =  \sqrt{2}
\Pi_1^{(1)}(u,u,b)~, \\
\label{ehv121111}
\Pi^{\Sigma_b^- \rar \Sigma_b^- \rho^0} =  \sqrt{2}
\Pi_1^{(1)}(d,d,b)~~.
\eea

The result for the
invariant function responsible for the $\Xi_b^{'-(0)} \rar \Xi_b^{'-(0)} \rho^0$
transitions can easily be obtained from the result for the $\Sigma_b^0 \rar
\Sigma_b^0 \rho^0$ transition using the fact that $\Xi_b^{'0} = \Sigma_b^0 (d
\rar s)$ and $\Xi_b^{'-} = \Sigma_b^0 (u \rar s)$, i.e.,
\bea
\label{ehv13}
\Pi^{\Xi_b^{'0} \rar \Xi_b^{'0} \rho^0} \es {1\over \sqrt{2}}
\Pi_1^{(1)} (u,s,b) ~, \\
\label{ehv14}
\Pi^{\Xi_b^{'-} \rar \Xi_b^{'-} \rho^0} \es- {1\over \sqrt{2}}
\Pi_1^{(1)} (d,s,b) ~.
\eea 

How can one find the relations among the invariant functions in
the presence of the charged $\rho^\pm$ mesons? In order to answer this 
question, we again consider the matrix element, $\lla \bar{d}d \vel \Sigma_b^0
\bar{\Sigma}_b^0 \ver 0 \rra$. This matrix element means that the $d$ quarks
from $\Sigma_b^0$ and $\bar{\Sigma}_b^0$ form the final $\bar{d}d$ state,
while $u$  and $b$ quarks are the spectators. The matrix element $\lla
\bar{u}d \vel \Sigma_b^+ \bar{\Sigma}_b^0 \ver 0 \rra$ corresponds to the
case when $d$ quark from $\bar{\Sigma}_b^0$ and $u$ quark from
$\Sigma_b^+$ form the $\bar{u}d$ state, again the remaining $u$ and $b$
quarks  being the spectators. This fact allows us to comment that these
matrix elements are proportional to each other. A detailed calculation shows
that,  these matrix elements are related to each other through,
\bea
\label{ehv15}
\Pi^{\Sigma_b^0 \rar \Sigma_b^+ \rho^-} \es \lla \bar{u}d \vel \Sigma_b^+
\bar{\Sigma}_b^0 \ver 0 \rra = -\sqrt{2} \lla \bar{d}d \vel \Sigma_b^0
\bar{\Sigma}_b^0 \ver 0 \rra \nnb \\
\es -\sqrt{2} \Pi_1^{(1)}(d,u,b)~.
\eea
Making the replacement $u \lrar d$, from Eq. (\ref{ehv15}) we get,
\bea
\label{ehv16}
\Pi^{\Sigma_b^0 \rar \Sigma_b^- \rho^+} = \sqrt{2} \Pi_1^{(1)}(u,d,b)~.
\eea

In calculating the coupling constants of SSV, SAV and AAV, it is enough to
consider the transitions $\Sigma_b^0 \rar \Sigma_b^0 V$,
$\Xi_b^{'0} \rar \Xi_b^{'0} V$ and $\Xi_b^0 \rar \Xi_b^0 V$, respectively,
since all other strong couplings can be achieved from these results with the
help of corresponding replacements among the quarks. 
In order to obtain the correlation functions in terms of $\Pi_1$,
which describe transitions among sextet--sextet, sextet--antitriplet and 
antitriplet--antitriplet with other vector mesons, similar calculations can
be done. These results are presented in Appendix A.

We can now proceed to calculate the invariant functions
$\Pi_1^{(i)}~(i=1,2,3)$  responsible for $\Sigma_b^0 \rar \Sigma_b^0 \rho^0$,
$\Xi_b^{'0} \rar \Xi_b^{'0} \rho^0$ and $\Xi_b^0 \rar \Xi_b^0
\rho^0$transitions from QCD side. These invariant functions can be
calculated in deep Eucledian region $-p^2 \rar \infty$, $-(p+q)^2 \rar
\infty$ using the operator product expansion. The main nonperturbative
ingredient in the calculations are the distribution amplitude (DA's) of the
vector mesons. These distribution amplitudes appear in determination of the
matrix elements of nonlocal operators $\lla V(q) \vel \bar{q} (x) \Gamma
q(0) \ver 0 \rra$ and $\lla V(q) \vel \bar{q} (x) G_{\mu\nu} q(0) \ver 0 
\rra$, where $\Gamma$ is any Dirac matrix. Up to twist twist--4 accuracy, the
expressions for the distribution functions of  vector mesons can be found in
\cite{Rhv11,Rhv12}.

In calculating the correlation functions responsible for afore--mentioned
decays from QCD side, the expressions of light and heavy quark propagators
are also needed.   
The light quark propagator in an external field is calculated in
\cite{Rhv13} whose expression is given as
\bea
\label{ehv17}
S_q(x) \es {i \rlap/x\over 2\pi^2 x^4} - {m_q\over 4 \pi^2 x^2} -
{\lla \bar q q \rra\over 12} \left(1 - i {m_q\over 4} \rlap/x \right) -
{x^2\over 192} m_0^2 \lla \bar q q \rra  \left( 1 -
i {m_q\over 6}\rlap/x \right) \nnb \\
&&  - i g_s \int_0^1 du \left[{\rlap/x\over 16 \pi^2 x^2} G_{\mu \nu} (ux)
\sigma_{\mu \nu} - {i\over 4 \pi^2 x^2} u x^\mu G_{\mu \nu} (ux) \gamma^\nu
\right. \nnb \\
&& \left.
 - i {m_q\over 32 \pi^2} G_{\mu \nu} \sigma^{\mu
 \nu} \left( \ln \left( {-x^2 \Lambda^2\over 4} \right) +
 2 \gamma_E \right) \right]~,
\eea
where $\gamma_E \simeq 0.577$ is the Euler Constant, and $\Lambda$ is the
scale parameter and it is chosen as the factorization scale 
$\Lambda=(0.5 \div 1)~GeV$ (for more detail see \cite{Rhv14}).

The propagator for the heavy quark is \cite{Rhv15}:
\bea
\label{ehv18}
S_Q(x) \es {m_Q^2 \over 4 \pi^2} {K_1(m_Q\sqrt{-x^2}) \over \sqrt{-x^2}} -
i {m_Q^2 \rlap/{x} \over 4 \pi^2 x^2} K_2(m_Q\sqrt{-x^2}) \nnb \\
\ek ig_s \int {d^4k \over (2\pi)^4} e^{-ikx} \int_0^1
du \Bigg[ {\rlap/k+m_Q \over 2 (m_Q^2-k^2)^2} G^{\mu\nu} (ux)
\sigma_{\mu\nu} +
{u \over m_Q^2-k^2} x_\mu G^{\mu\nu} \gamma_\nu \Bigg]~,
\eea
where
$K_i$ are the modified Bessel function of the second kind.

Using the expressions of light and heavy quark propagators, as well as
definitions of the DA's for the vector mesons, and after lengthy
calculations one can obtain the correlation function from QCD part.

As has already been noted, the relations among the correlation functions for
the considered transitions are structure independent, but their explicit
expressions are structure dependent. For this reason we introduce new
indices $\alpha$ in the correlation function, where $\alpha=1$ stands for
the choice of the structure $(\varepsilon\!\cdot \!p) \rlap/{p}$ and
$\alpha=2$ for the structure $\rlap/{p}\rlap/{\varepsilon}\rlap/{q}$.

Equating the coefficients of the structures, $(\varepsilon\!\cdot \!p)
\rlap/{p}$ and $\rlap/{p}\rlap/{\varepsilon}\rlap/{q}$ for the hadronic and
QCD sides and performing Borel transformations over the
variables $p^2$ and $(p+q)^2$, which suppress the contributions of the
continuum and higher states, we finally get the sum rules for the strong
coupling constants of light vector mesons with sextet and antitriplet heavy
baryons as:
\bea
\label{ehv19}f^{(i)}_\alpha = {1 \over \lambda_1^{(i)} \lambda_2^{(i)}} e^{{m_1^{(i)2}
\over M_1^2} + {m_2^{(i)2}
\over M_2^2} + {m_V^2 \over M_1^2 + M_2^2}}\, \Pi_\alpha^{(i)}~,
\eea
where $M_1^2$ and $M_2^2$ are the Borel parameters corresponding to the
initial and final heavy baryons, respectively.
In the problem under consideration the masses of the initial and final heavy
baryons are very close to each other, hence we can take $M_1^2=M_2^2\equiv2
M^2$. Note that the residues of the sextet and antitriplet heavy baryons are
calculated in \cite{Rhv16}. This leads to the result that for numerical
calculation of the sum rules, the DA's are needed to be evaluated only at
$u_0={M_1^2 \over M_1^2+M_2^2} = {1 \over2}$.

\section{Numerical analysis}

Having already obtained the sum rules for the strong coupling constants of
light vector mesons with heavy baryons, we can now proceed evaluating them
numerically. The essential ingredients of the LCSR are the  DA's
of the light vector mesons. Explicit expressions of DA's for the vector
mesons and the values of the parameters entering to the expressions of DA's
are given in \cite{Rhv11,Rhv12}. The residues of the heavy baryons are
calculated in \cite{Rhv16}.

In addition to the DA's and other input parameters, the sum rules for SSV, SAV and AAV transitions contain three auxiliary
parameters : the continuum threshold $s_0$, Borel
parameter $M^2$ and the parameter $\beta$ entering  the expressions of
the interpolating current. For this reason we try to find such regions of
these parameters where strong coupling constants are practically independent
of them.

In finding the working region of $M^2$, we require that the continuum and
higher state contributions should be less than half of the dispersion
integral, and additionally the contribution of the higher terms with the
power $1/M^2$ be 25\% less than the total result.
These two restrictions lead to the result that the ``working region" of
$M^2$ is $15~GeV^2 \le M^2 \le 30~GeV^2$ (for the bottom baryons) and
$4~GeV^2 \le M^2 \le 12~GeV^2$ (for the charmed ones). The continuum threshold
is varied in the region $(m_B+0.5)^2~GeV^2 \le s_0 \le (m_B+0.7)^2~GeV^2$.
To be more illustrative about how the numerical analysis is performed, as an
example, we consider the $\Xi_b^{'0} \rar \Xi_b^{'0} \rho^0$ transition. In
Figs. (1) and (2) we present the dependence of $f_1$ and $f_2$ on $M^2$
for the above--mentioned transition, at five different values of $\beta$,
and at a fixed value of $s_0$. From these figures, one can conclude that the strong coupling constants, $f_1$ and $f_2$ are practically
independent of $M^2$ when it varies in its own `working region".    

In Figs. (3) and (4) we present the dependence of $f_1$ and $f_2$ on
$\cos\theta$, at three fixed values of $s_0$ and 
$M^2$, where $\tan\theta=\beta$. We
observe from these figures that when $\cos\theta$ is varied in the region
$-0.5 \le \cos\theta\le 0.3$, the results are insensitive to the variation
of $\beta$. From these figures, we obtain, $f_1^{\Xi_b^{'0} \rar
\Xi_b^{'0} \rho^0}=2.2 \pm 0.7$ and $f_2^{\Xi_b^{'0} \rar \Xi_b^{'0} \rho^0}=
30 \pm 10$.
Similar analysis for the other couplings of vector mesons are carried out
and the results for $f_1$ and $f_2$ are presented in Tables (2)--(7), 
respectively. The errors presented in these Tables are due to the variation of
the auxiliary parameters, as well as uncertainties in the values  of the
input parameters.

% .........................................................

\begin{table}[t]

\renewcommand{\arraystretch}{1.3}
\addtolength{\arraycolsep}{-0.5pt}
\small
$$
\begin{array}{|l|r@{\pm}l|r@{\pm}l||l|r@{\pm}l|r@{\pm}l|}
\hline \hline  
 \multirow{2}{*}{$f_1^{\mbox{\small{\,channel}}}$}        &\multicolumn{4}{c||}{\mbox{Bottom Baryons}}   &  
 \multirow{2}{*}{$f_1^{\mbox{\small{\,channel}}}$}        &\multicolumn{4}{c|}{\mbox{Charmed Baryons}} \\
	                                                &   \multicolumn{2}{c}{\mbox{~General current~}}        & 
	                                                    \multicolumn{2}{c||}{\mbox{~Ioffe current~}}       & &
                                                                                                                   \multicolumn{2}{|c}{\mbox{~General current~}}  & 
                                                                                                                   \multicolumn{2}{c|}{\mbox{~Ioffe current~}}       \\ \hline
 f_1^{\Xi_b^{'0}    \rar \Xi_b^{'0}    \rho^0}            &~~~~~~~2.2&0.7    &~~~~~2.0&0.7 &
 f_1^{\Xi_c^{'+}    \rar \Xi_c^{'+}    \rho^0}            &~~~~~~ 2.5&0.8    &~~~~ 5.0&1.7   \\ 
 f_1^{\Sigma_b^0    \rar \Sigma_b^-    \rho^+}            &       4.5&1.5    &    3.4&1.1  &
 f_1^{\Sigma_c^+    \rar \Sigma_c^0    \rho^+}            &       4.0&1.3    &    3.4&1.1   \\
 f_1^{\Xi_b^{'0}    \rar \Sigma_b^+     K^{\ast-}}        &       6.0&2.0    &    3.9&1.3  &
 f_1^{\Xi_c^{'+}    \rar \Sigma_c^{++}  K^{\ast-}}        &       5.0&1.7    &    3.8&1.3   \\
 f_1^{\Omega_b^-    \rar \Xi_b^{'0}     \bar{K}^{\ast-}}  &       6.0&2.0    &    4.8&1.6  &
 f_1^{\Omega_c^0    \rar \Xi_c^{'+}     \bar{K}^{\ast-}}  &       7.0&2.0    &   14.0&5.0   \\
 f_1^{\Sigma_b^+    \rar \Sigma_b^+    \omega}            &       4.0&1.3    &    3.0&1.0  &
 f_1^{\Sigma_c^{++} \rar \Sigma_c^{++} \omega}            &       3.5&1.2    &    3.0&1.0   \\
 f_1^{\Xi_b^{'0}    \rar \Xi_b^{'0}    \omega}            &       2.1&0.7    &    1.7&0.6  &
 f_1^{\Xi_c^{'+}    \rar \Xi_c^{'+}    \omega}            &       2.4&0.8    &    4.9&1.6   \\
 f_1^{\Xi_b^{'0}    \rar \Xi_b^{'0}    \phi}              &       5.0&1.7    &    2.6&0.9  &
 f_1^{\Xi_c^{'+}    \rar \Xi_c^{'+}    \phi}              &       4.0&1.3    &    2.5&0.8   \\
 f_1^{\Omega_b^-    \rar \Omega_b^-    \phi}              &      10.0&3.4    &    7.0&2.4  &
 f_1^{\Omega_c^0    \rar \Omega_c^0    \phi}              &      11.0&4.0    &   23.0&8.0   \\

 \hline \hline
\end{array}
$$

\caption{The values of the strong coupling constants $f_1$ for the transitions
among the sextet--sextet heavy baryons with vector mesons.}

\renewcommand{\arraystretch}{1}
\addtolength{\arraycolsep}{-1.0pt}

\end{table}

% .........................................................

\begin{table}[h]

\renewcommand{\arraystretch}{1.3}
\addtolength{\arraycolsep}{-0.5pt}
\small
$$
\begin{array}{|l|r@{\pm}l|r@{\pm}l||l|r@{\pm}l|r@{\pm}l|}
\hline \hline  
 \multirow{2}{*}{$f_1^{\mbox{\small{\,channel}}}$}        &\multicolumn{4}{c||}{\mbox{Bottom Baryons}}   &  
 \multirow{2}{*}{$f_1^{\mbox{\small{\,channel}}}$}        &\multicolumn{4}{c|}{\mbox{Charmed Baryons}} \\
	                                                &   \multicolumn{2}{c}{\mbox{~General current~}}        & 
	                                                    \multicolumn{2}{c||}{\mbox{~Ioffe current~}}       & &
                                                                                                                   \multicolumn{2}{|c}{\mbox{~General current~}}  & 
                                                                                                                   \multicolumn{2}{c|}{\mbox{~Ioffe current~}}       \\ \hline
 f_1^{\Xi_b^{'0}    \rar \Xi_b^0       \rho^0}          &~~~~~~~1.4&0.5    &~~~~~0.6&0.2   &
 f_1^{\Xi_c^{'+}    \rar \Xi_c^+       \rho^0}          &~~~~~~ 1.5&0.5    &~~~~ 0.7&0.2   \\
 f_1^{\Xi_b^{'0}    \rar \Xi_b^-        K^{\ast+}}      &       2.5&0.8    &     1.3&0.4  &
 f_1^{\Xi_c^{'+}    \rar \Xi_c^0        K^{\ast+}}      &       2.3&0.8    &     1.2&0.4   \\
 f_1^{\Sigma_b^-    \rar \Lambda_b^0   \rho^-}          &       2.8&0.9    &     0.8&0.3  &
 f_1^{\Sigma_c^0    \rar \Lambda_c^+   \rho^-}          &       2.6&0.9    &     0.6&0.2   \\
 f_1^{\Sigma_b^0    \rar \Xi_b^0       \bar{K}^{\ast0}} &       2.6&0.8    &     1.5&0.5  &
 f_1^{\Sigma_c^+    \rar \Xi_c^+       \bar{K}^{\ast0}} &       2.2&0.7    &     0.8&0.3   \\
 f_1^{\Omega_b^-    \rar \Xi_b^-       \bar{K}^{\ast0}} &       3.5&1.2    &     2.0&0.6  &
 f_1^{\Omega_c^0    \rar \Xi_c^0       \bar{K}^{\ast0}} &       3.3&1.1    &     1.7&0.6   \\
 f_1^{\Xi_b^{'0}    \rar \Xi_b^-       \omega}          &       1.3&0.4    &     0.5&0.2  &
 f_1^{\Xi_c^{'+}    \rar \Xi_c^0       \omega}          &       1.2&0.4    &     1.1&0.4   \\
 f_1^{\Xi_b^{'0}    \rar \Xi_b^0       \phi}            &       2.6&0.9    &     2.0&0.7  &
 f_1^{\Xi_c^{'+}    \rar \Xi_c^+       \phi}            &       2.1&0.7    &     1.4&0.5   \\
 \hline \hline
\end{array}
$$

\caption{The values of the strong coupling constants $f_1$ for the transitions
among the sextet--antitriplet heavy baryons with vector mesons.}

\renewcommand{\arraystretch}{1}
\addtolength{\arraycolsep}{-1.0pt}

\end{table}
% .........................................................

\begin{table}[h]

\renewcommand{\arraystretch}{1.3}
\addtolength{\arraycolsep}{-0.5pt}
\small
$$
\begin{array}{|l|r@{\pm}l|r@{\pm}l||l|r@{\pm}l|r@{\pm}l|}
\hline \hline  
 \multirow{2}{*}{$f_1^{\mbox{\small{\,channel}}}$}        &\multicolumn{4}{c||}{\mbox{Bottom Baryons}}   &  
 \multirow{2}{*}{$f_1^{\mbox{\small{\,channel}}}$}        &\multicolumn{4}{c|}{\mbox{Charmed Baryons}} \\
	                                                &   \multicolumn{2}{c}{\mbox{~General current~}} & 
	                                                    \multicolumn{2}{c||}{\mbox{~Ioffe current~}} & &
                                                            \multicolumn{2}{|c}{\mbox{~General current~}}  & 
                                                            \multicolumn{2}{c|}{\mbox{~Ioffe current~}} \\ \hline
 f_1^{\Xi_b^0       \rar \Xi_b^0       \rho^0}          &~~~~~~~3.1&1.1    &~~~~2.5&0.8  &
 f_1^{\Xi_c^+       \rar \Xi_c^+       \rho^0}          &~~~~~~~6.0&2.0    &~~~~1.5&0.5   \\
 f_1^{\Xi_b^-       \rar \Lambda_b^0    K^{\ast-}}      &       5.0&1.7    &     4.6&1.5  &
 f_1^{\Xi_c^0       \rar \Lambda_c^+    K^{\ast-}}      &       4.6&1.5    &     4.1&1.4   \\
 f_1^{\Xi_b^0       \rar \Xi_b^0       \omega}          &       2.8&0.9    &     2.3&0.8  &
 f_1^{\Xi_c^+       \rar \Xi_c^+       \omega}          &       5.5&1.8    &     1.2&0.4   \\
 f_1^{\Lambda_b^0   \rar \Lambda_b^0   \omega}          &       5.2&1.7    &     4.6&1.5  &
 f_1^{\Lambda_c^+   \rar \Lambda_c^+   \omega}          &       4.9&1.6    &     4.3&1.4   \\
 f_1^{\Lambda_b^0   \rar \Lambda_b^0   \phi}            &       5.0&1.7    &     4.7&1.6  &
 f_1^{\Lambda_c^+   \rar \Lambda_c^+   \phi}            &       4.6&1.5    &     4.1&1.4  \\
 \hline \hline
\end{array}
$$

\caption{The values of the strong coupling constants $f_1$ for the transitions
among the antitriplet--antitriplet heavy baryons with vector mesons.}

\renewcommand{\arraystretch}{1}
\addtolength{\arraycolsep}{-1.0pt}

\end{table}

% .........................................................

\begin{table}[t]

\renewcommand{\arraystretch}{1.3}
\addtolength{\arraycolsep}{-0.5pt}
\small
$$
\begin{array}{|l|r@{\pm}l|r@{\pm}l||l|r@{\pm}l|r@{\pm}l|}
\hline \hline  
 \multirow{2}{*}{$f_2^{\mbox{\small{\,channel}}}$}        &\multicolumn{4}{c||}{\mbox{Bottom Baryons}}    &  
 \multirow{2}{*}{$f_2^{\mbox{\small{\,channel}}}$}        &\multicolumn{4}{c|}{\mbox{Charmed Baryons}} \\
	                                                &   \multicolumn{2}{c}{\mbox{~General current~}}  & 
	                                                    \multicolumn{2}{c||}{\mbox{~Ioffe current~}}  & &
                                                            \multicolumn{2}{|c}{\mbox{~General current~}} & 
                                                            \multicolumn{2}{c|}{\mbox{~Ioffe current~}} \\ \hline
 f_2^{\Xi_b^{'0}    \rar \Xi_b^{'0}    \rho^0}            &~~~~~30.0&10.0~~~&~~~~~39.0&13.0~~~~  &
 f_2^{\Xi_c^{'+}    \rar \Xi_c^{'+}    \rho^0}            &~~~~~16.0&5.2 ~~~&~~~~~18.0&6.0 ~~~~   \\ 
 f_2^{\Sigma_b^0    \rar \Sigma_b^-    \rho^+}            &       55.0&18.0  &     72.0&24.0     &
 f_2^{\Sigma_c^+    \rar \Sigma_c^0    \rho^+}            &       27.0&9.0   &     33.0&11.0      \\
 f_2^{\Xi_b^{'0}    \rar \Sigma_b^+     K^{\ast-}}        &       60.0&20.0  &     80.0&27.0     &
 f_2^{\Xi_c^{'+}    \rar \Sigma_c^{++}  K^{\ast-}}        &       30.0&10.0  &     36.0&12.0      \\
 f_2^{\Omega_b^-    \rar \Xi_b^{'0}     \bar{K}^{\ast-}}  &       70.0&23.0  &     88.0&29.0     &
 f_2^{\Omega_c^0    \rar \Xi_c^{'+}     \bar{K}^{\ast-}}  &       35.0&12.0  &     41.0&14.0      \\
 f_2^{\Sigma_b^+    \rar \Sigma_b^+    \omega}            &       50.0&17.0  &     64.0&21.0     &
 f_2^{\Sigma_c^{++} \rar \Sigma_c^{++} \omega}            &       24.0&8.0   &     29.0&9.5       \\
 f_2^{\Xi_b^{'0}    \rar \Xi_b^{'0}    \omega}            &       27.0&9.0   &     34.0&11.0     &
 f_2^{\Xi_c^{'+}    \rar \Xi_c^{'+}    \omega}            &       15.0&5.0   &     16.0&6.3       \\
 f_2^{\Xi_b^{'0}    \rar \Xi_b^{'0}    \phi}              &       45.0&15.0  &     57.0&19.0     &
 f_2^{\Xi_c^{'+}    \rar \Xi_c^{'+}    \phi}              &       21.0&7.0   &     26.0&8.6       \\
 f_2^{\Omega_b^-    \rar \Omega_b^-    \phi}              &       95.0&32.0  &    125.0&42.0     &
 f_2^{\Omega_c^0    \rar \Omega_c^0    \phi}              &       52.0&17.0    &   60.0&20.0      \\
 \hline \hline
\end{array}
$$

\caption{The values of the strong coupling constants $f_2$ for the transitions
among the sextet--sextet heavy baryons with vector mesons.}

\renewcommand{\arraystretch}{1}
\addtolength{\arraycolsep}{-1.0pt}

\end{table}

% .........................................................

\begin{table}[t]

\renewcommand{\arraystretch}{1.3}
\addtolength{\arraycolsep}{-0.5pt}
\small
$$
\begin{array}{|l|r@{\pm}l|r@{\pm}l||l|r@{\pm}l|r@{\pm}l|}
\hline \hline  
 \multirow{2}{*}{$f_2^{\mbox{\small{\,channel}}}$}        &\multicolumn{4}{c||}{\mbox{Bottom Baryons}}    &  
 \multirow{2}{*}{$f_2^{\mbox{\small{\,channel}}}$}        &\multicolumn{4}{c|}{\mbox{Charmed Baryons}} \\
	                                                &   \multicolumn{2}{c}{\mbox{~General current~}}  & 
	                                                    \multicolumn{2}{c||}{\mbox{~Ioffe current~}}  & &
                                                            \multicolumn{2}{|c}{\mbox{~General current~}} & 
                                                            \multicolumn{2}{c|}{\mbox{~Ioffe current~}} \\ \hline
 f_2^{\Xi_b^{'0}    \rar \Xi_b^0       \rho^0}          &~~~~~22.0&7.0~~~  &~~~~23.0&7.0~~~~&
 f_2^{\Xi_c^{'+}    \rar \Xi_c^+       \rho^0}          &~~~~~~11.0&3.8~~~ &~~~~11.0&3.8~~~~ \\
 f_2^{\Xi_b^{'0}    \rar \Xi_b^-        K^{\ast+}}      &      34.0&11.0   &    38.0&13.0   &
 f_2^{\Xi_c^{'+}    \rar \Xi_c^0        K^{\ast+}}      &      15.0&5.0    &    18.0&6.0     \\
 f_2^{\Sigma_b^-    \rar \Lambda_b^0   \rho^-}          &      40.0&13.0   &    42.0&14.0   &
 f_2^{\Sigma_c^0    \rar \Lambda_c^+   \rho^-}          &      16.0&5.3    &    18.0&6.0     \\
 f_2^{\Sigma_b^0    \rar \Xi_b^0       \bar{K}^{\ast0}} &      32.0&11.0   &    35.0&12.0   &
 f_2^{\Sigma_c^+    \rar \Xi_c^+       \bar{K}^{\ast0}} &      13.0&4.3    &    15.0&5.0     \\
 f_2^{\Omega_b^-    \rar \Xi_b^-       \bar{K}^0}       &      50.0&17.0   &    53.0&18.0   &
 f_2^{\Omega_c^0    \rar \Xi_c^0       \bar{K}^0}       &      20.0&7.0    &    26.0&8.5     \\
 f_2^{\Xi_b^{'0}    \rar \Xi_b^-       \omega}          &      20.3&7.0    &    20.0&7.0    &
 f_2^{\Xi_c^{'+}    \rar \Xi_c^0       \omega}          &       8.0&2.7    &    10.0&3.0     \\
 f_2^{\Xi_b^{'0}    \rar \Xi_b^0       \phi}            &      30.0&10.0   &    34.0&11.0   &
 f_2^{\Xi_c^{'+}    \rar \Xi_c^+       \phi}            &      13.0&4.3    &    15.0&5.0     \\
 \hline \hline
\end{array}
$$

\caption{The values of the strong coupling constants $f_2$ for the transitions
among the sextet--antitriplet heavy baryons with vector mesons.}

\renewcommand{\arraystretch}{1}
\addtolength{\arraycolsep}{-1.0pt}

\end{table}
% .........................................................

\begin{table}[t]

\renewcommand{\arraystretch}{1.3}
\addtolength{\arraycolsep}{-0.5pt}
\small
$$
\begin{array}{|l|r@{\pm}l|r@{\pm}l||l|r@{\pm}l|r@{\pm}l|}
\hline \hline  
 \multirow{2}{*}{$f_2^{\mbox{\small{\,channel}}}$}        &\multicolumn{4}{c||}{\mbox{Bottom Baryons}}   &  
 \multirow{2}{*}{$f_2^{\mbox{\small{\,channel}}}$}        &\multicolumn{4}{c|}{\mbox{Charmed Baryons}} \\
	                                                &   \multicolumn{2}{c}{\mbox{~General current~}} & 
	                                                    \multicolumn{2}{c||}{\mbox{~Ioffe current~}} & &
                                                            \multicolumn{2}{|c}{\mbox{~General current~}}  & 
                                                            \multicolumn{2}{c|}{\mbox{~Ioffe current~}} \\ \hline
 f_2^{\Xi_b^0       \rar \Xi_b^0       \rho^0}          &~~~~~~~5.0&1.7    &~~~~12.0&3.8  &
 f_2^{\Xi_c^+       \rar \Xi_c^+       \rho^0}          &~~~~~~~7.5&2.5    &~~~~ 5.7&1.9   \\
 f_2^{\Xi_b^-       \rar \Lambda_b^0    K^{\ast-}}      &       7.0&2.0    &    24.0&8.0  &
 f_2^{\Xi_c^0       \rar \Lambda_c^+    K^{\ast-}}      &       6.0&2.0    &    12.0&4.0   \\
 f_2^{\Xi_b^0       \rar \Xi_b^0       \omega}          &       4.0&1.3    &    11.0&3.6  &
 f_2^{\Xi_c^+       \rar \Xi_c^+       \omega}          &       7.5&2.5    &     5.1&1.7   \\
 f_2^{\Lambda_b^0   \rar \Lambda_b^0   \omega}          &       8.0&2.7    &    25.0&8.0  &
 f_2^{\Lambda_c^+   \rar \Lambda_c^+   \omega}          &       6.0&2.0    &    14.0&5.0   \\
 f_2^{\Lambda_b^0   \rar \Lambda_b^0   \phi}            &       8.0&2.7    &    23.0&7.5  &
 f_2^{\Lambda_c^+   \rar \Lambda_c^+   \phi}            &       6.0&2.0    &    12.0&4.0  \\
 \hline \hline
\end{array}
$$

\caption{The values of the strong coupling constants $f_2$ for the transitions
among the antitriplet--antitriplet heavy baryons with vector mesons.}

\renewcommand{\arraystretch}{1}
\addtolength{\arraycolsep}{-1.0pt}

\end{table}

The results on the coupling constants of the light vector mesons with heavy
baryons  presented in Tables (2)--(7) lead to the following
conclusions:

\begin{itemize}
\item For the coupling constant, $f_1$: The predictions for the strong
coupling constant $f_1$ which are obtained using the most general and Ioffe
currents ($\beta=-1$) disagree considerably from each other, especially for
the channels, $\Xi_b^{'0} \rar \Xi_b^{'0} \phi$, $\Xi_c^{'+} \rar \Xi_c^{'+}
\omega$, $\Omega_c^{0} \rar \Xi_c^{'+} K^{\ast -}$, $\Omega_c^{0} \rar \Omega_c^{0} \phi$,
$\Xi_c^+ \rar \Xi_c^+ \rho^0$, $\Xi_c^{'+} \rar \Xi_c^+ \rho^0$, $\Xi_c^{'+} \rar \Xi_c^{'+} \rho^0$, $\Xi_b^{'0} \rar \Xi_b^{0} \rho^0$,
 $\Xi_c^{+} \rar \Xi_c^+ \omega$, $\Xi_b^{'0} \rar \Xi_b^- \omega$, 
$\Sigma_b^- \rar \Lambda_b^0 \rho^-$ and
$\Sigma_c^0 \rar \Lambda_c^+ \rho^-$.

\item For the $f_2$ channel: predictions of the general  and Ioffe
currents for the strong coupling constants of SSV and AAV show considerable differences. These discrepancies can be attributed to the fact that,
 $\beta=-1$ lies outside the stability region of $\beta$
and it makes the predictions less reliable.

Our final remark in this section is that the coupling constants of the $\Xi_c^0 \rar \Xi_c^0 \rho^0$ , $\Xi_c^0 \rar
\Xi_c^0 \omega$ and $\Xi_c^0 \rar \Xi_c^0 \phi$ transitions are also calculated
in the HQET \cite{Rhv17} within the same framework as our work. Our
predictions on the coupling constants of these transitions and those  given in \cite{Rhv17} are in a good agreement with each other.   
\end{itemize}   
\section{Acknowledgment}
The authors would like to thank to A. Ozpineci and V. S. Zamiralov for useful discussions.

\newpage

\bAPP{A}{}

Here in this appendix we present the expressions of the correlation functions
in terms of invariant function $\Pi_1^{(i)}$ involving 
$\rho$, $K$, $omega$ and $\phi$ mesons.

\begin{itemize}
\item Correlation functions responsible for the 
sextet--sextet transitions.
\end{itemize}
\baeeq
\label{nolabel}
%6
\Pi^{\Sigma_b^+ \rar \Sigma_b^0     \rho^+ } \es 
\sqrt{2} \Pi_1^{(1)}(d,u,b)~, \nnb \\
%7
\Pi^{\Sigma_b^0 \rar \Sigma_b^-     \rho^+ } \es 
\sqrt{2} \Pi_1^{(1)}(u,d,b)~, \nnb \\
%8
\Pi^{\Xi_b^{'0} \rar \Xi_b^{'-}     \rho^+ } \es 
\Pi_1^{(1)}(d,s,b)~, \nnb \\
%9
\Pi^{\Sigma_b^0 \rar \Sigma_b^+     \rho^- } \es
\sqrt{2} \Pi_1^{(1)}(d,u,b)~, \nnb \\
%10
\Pi^{\Sigma_b^- \rar \Sigma_b^0     \rho^- } \es 
\sqrt{2} \Pi_1^{(1)}(u,d,b)~, \nnb \\
%11
\Pi^{\Xi_b^{'-} \rar \Xi_b^{'0}     \rho^- } \es 
\Pi_1^{(1)}(u,s,b)~, \nnb \\
%12
\Pi^{\Xi_b^{'0} \rar \Sigma_b^+     K^{\ast-}} \es 
\sqrt{2} \Pi_1^{(1)}(u,u,b)~, \nnb \\
%13
\Pi^{\Xi_b^{'-} \rar \Sigma_b^0     K^{\ast-}} \es 
\Pi_1^{(1)}(u,d,b)~, \nnb \\
%14
\Pi^{\Omega_b^- \rar \Xi_b^{'0}     K^{\ast-}} \es 
\sqrt{2} \Pi_1^{(1)}(s,s,b)~, \nnb \\
%15
\Pi^{\Sigma_b^+ \rar \Xi_b^{'0}     K^{\ast+}} \es 
\sqrt{2} \Pi_1^{(1)}(u,u,b)~, \nnb \\
%16
\Pi^{\Sigma_b^0 \rar \Xi_b^{'-}     K^{\ast+}} \es 
\Pi_1^{(1)}(u,d,b)~, \nnb \\
%17
\Pi^{\Xi_b^{'0} \rar \Omega_b^-      K^{\ast+}} \es 
\sqrt{2}\Pi_1^{(1)}(s,s,b)~, \nnb \\
%18
\Pi^{\Xi_b^{'0} \rar \Sigma_b^0     \bar{K}^{\ast0}} \es 
\Pi_1^{(1)}(d,u,b)~, \nnb \\
%19
\Pi^{\Xi_b^{'-} \rar \Sigma_b^-     \bar{K}^{\ast0}} \es 
\sqrt{2} \Pi_1^{(1)}(d,d,b)~, \nnb \\
%20
\Pi^{\Omega_b^-  \rar \Xi_b^{'-}    \bar{K}^{\ast0}} \es 
\sqrt{2} \Pi_1^{(1)}(s,s,b)~, \nnb \\
%21
\Pi^{\Sigma_b^0 \rar \Xi_b^{'0}     K^{\ast0}} \es 
\Pi_1^{(1)}(d,u,b)~, \nnb \\
%22
\Pi^{\Sigma_b^- \rar \Xi_b^{'-}     K^{\ast0}} \es 
\sqrt{2} \Pi_1^{(1)}(d,d,b)~, \nnb \\
%23
\Pi^{\Xi_b^{'-} \rar \Omega_b^-     K^{\ast0}} \es 
\sqrt{2} \Pi_1^{(1)}(s,s,b)~, \nnb \\
%24
\Pi^{\Sigma_b^0 \rar \Sigma_b^0    \omega} \es 
{1\over \sqrt{2}}\Big[ \Pi_1^{(1)}(u,d,b) + \Pi_1^{(1)}(d,u,b)\Big]~, \nnb \\
%25
\Pi^{\Sigma_b^+ \rar \Sigma_b^+    \omega} \es
\sqrt{2} \Pi_1^{(1)}(u,u,b)~, \nnb \\
%26
\Pi^{\Sigma_b^- \rar \Sigma_b^-    \omega} \es 
\sqrt{2} \Pi_1^{(1)}(d,d,b)~, \nnb \\
%27
\Pi^{\Xi_b^{'0} \rar \Xi_b^{'0}     \omega } \es
{1\over \sqrt{2}} \Pi_1^{(1)}(u,s,b)~, \nnb \\
%28
\Pi^{\Xi_b^{'-} \rar \Xi_b^{'-}     \omega } \es
{1\over \sqrt{2}} \Pi_1^{(1)}(d,s,b)~, \nnb \\
%33
\Pi^{\Xi_b^{'0} \rar \Xi_b^{'0}     \phi } \es
\Pi_1^{(1)}(s,u,b)~, \nnb \\
%34
\Pi^{\Xi_b^{'-} \rar \Xi_b^{'-}     \phi } \es
\Pi_1^{(1)}(s,d,b)~, \nnb \\
%35
\Pi^{\Omega_b^- \rar \Omega_b^-     \phi } \es
2 \Pi_1^{(1)}(s,s,b)~.~~~~~~~~~~~~~~~~~~~~~~~~~~~~~~~
\eaeeq

\begin{itemize}
\item Correlation functions responsible for the 
sextet--antitriplet transitions.
\end{itemize}
\baeeq
\label{nolabel}
%36
\Pi^{\Xi_b^{'0} \rar \Xi_b^0     \rho^0 } \es
{1\over \sqrt{2}} \Pi_2^{(1)}(u,s,b)~, \nnb \\
%37
\Pi^{\Xi_b^{'-} \rar \Xi_b^-     \rho^0 } \es
- {1\over \sqrt{2}} \Pi_2^{(1)}(d,s,b)~, \nnb \\
%38
\Pi^{\Sigma_b^0 \rar \Lambda_b     \rho^0 } \es
{1\over \sqrt{2}} \Big[\Pi_2^{(1)}(u,d,b) - \Pi_2^{(1)}(d,u,b) \Big]~, \nnb \\
%39
\Pi^{\Sigma_b^- \rar \Lambda_b     \rho^- } \es
\sqrt{2} \Pi_2^{(1)}(u,d,b)~, \nnb \\
%40
\Pi^{\Xi_b^{'-} \rar \Xi_b^-      \rho^- } \es
\Pi_2^{(1)}(d,s,b)~, \nnb \\
%41
\Pi^{\Sigma_b^+ \rar \Lambda_b     \rho^+ } \es
- \sqrt{2} \Pi_2^{(1)}(d,u,b)~, \nnb \\
%42
\Pi^{\Xi_b^{'0} \rar \Xi_b^-      \rho^+ } \es
\Pi_2^{(1)}(u,s,b)~, \nnb \\
%43
\Pi^{\Sigma_b^0 \rar \Xi_b^0      \bar{K}^{\ast0}} \es
- \Pi_2^{(1)}(d,u,b)~, \nnb \\
%44
\Pi^{\Sigma_b^- \rar \Xi_b^-      \bar{K}^{\ast0}} \es
- \sqrt{2} \Pi_2^{(1)}(d,d,b)~, \nnb \\
%45
\Pi^{\Omega_b^- \rar \Xi_b^-      \bar{K}^{\ast0}} \es
\sqrt{2} \Pi_2^{(1)}(s,s,b)~, \nnb \\ 
%46
\Pi^{\Xi_b^{'0} \rar \Lambda_b    \bar{K}^{\ast0}} \es
- \Pi_2^{(1)}(d,u,b)~, \nnb \\
%47
\Pi^{\Sigma_b^0 \rar \Xi_b^0           K^{\ast0}} \es
- \Pi_2^{(1)}(d,u,b)~, \nnb \\
%48
\Pi^{\Sigma_b^- \rar \Xi_b^-           K^{\ast0}} \es
- \sqrt{2} \Pi_2^{(1)}(d,d,b)~, \nnb \\
%49
\Pi^{\Omega_b^- \rar \Xi_b^-           K^{\ast0}} \es
\sqrt{2} \Pi_2^{(1)}(s,s,b)~, \nnb \\ 
%50
\Pi^{\Xi_b^{'0} \rar \Lambda_b         K^{\ast0}} \es
- \Pi_2^{(1)}(d,u,b)~, \nnb \\
%51
\Pi^{\Sigma_b^+ \rar \Lambda_b         K^{\ast+}} \es
- \sqrt{2} \Pi_2^{(1)}(u,u,b)~, \nnb \\
%52
\Pi^{\Sigma_b^0 \rar \Xi_b^-           K^{\ast+}} \es
- \Pi_2^{(1)}(u,d,b)~, \nnb \\
%53
\Pi^{\Xi_b^{'0} \rar \Xi_b^-           K^{\ast+}} \es
\Pi_2^{(1)}(d,s,b)~, \nnb \\
%54
\Pi^{\Sigma_b^- \rar \Lambda_b         K^{\ast-}} \es
\sqrt{2} \Pi_2^{(1)}(d,d,b)~, \nnb \\
%55
\Pi^{\Omega_b^- \rar \Xi_b^0           K^{\ast-}} \es
\sqrt{2} \Pi_2^{(1)}(s,s,b)~, \nnb \\ 
%56
\Pi^{\Xi_b^{'-} \rar \Xi_b^0           K^{\ast-}} \es
\Pi_2^{(1)}(u,s,b)~, \nnb \\
%57
\Pi^{\Xi_b^{'0} \rar \Xi_b^0          \omega}     \es
{1\over \sqrt{2}} \Pi_2^{(1)}(d,s,b)~, \nnb \\
%58
\Pi^{\Xi_b^{'-} \rar \Xi_b^-          \omega}     \es    
{1\over \sqrt{2}} \Pi_2^{(1)}(d,s,b)~, \nnb \\    
%59
\Pi^{\Sigma_b^0 \rar \Lambda_b        \omega}     \es
{1\over \sqrt{2}} \Big[\Pi_2^{(1)}(u,d,b) - \Pi_2^{(1)}(d,u,b) \Big]~, \nnb \\
%60
\Pi^{\Xi_b^{'0} \rar \Xi_b^0          \phi}       \es
- \Pi_2^{(1)}(s,u,b)~, \nnb \\
%61
\Pi^{\Xi_b^{'-} \rar \Xi_b^-          \phi}       \es
- \Pi_2^{(1)}(s,d,b)~.
\eaeeq

\begin{itemize}
\item Correlation functions responsible for the
antitiplet--antitriplet transitions.
\end{itemize}
\baeeq
\label{nolabel}
%63
\Pi^{\Xi_b^0 \rar \Xi_b^0             \rho^0}    \es
{1\over \sqrt{2}} \Pi_3^{(1)}(u,s,b)~, \nnb \\
%64
\Pi^{\Xi_b^- \rar \Xi_b^-             \rho^0}    \es
- {1\over \sqrt{2}} \Pi_3^{(1)}(d,s,b)~, \nnb \\
%65
\Pi^{\Lambda_b \rar \Lambda_b         \rho^0}    \es
- {1\over \sqrt{2}} \Big[\Pi_3^{(1)}(d,u,b) - \Pi_3^{(1)}(u,d,b) \Big]~, \nnb \\
%66
\Pi^{\Xi_b^- \rar \Xi_b^0             \rho^-}    \es
\Pi_3^{(1)}(d,s,b)~, \nnb \\
%67
\Pi^{\Xi_b^0 \rar \Xi_b^-             \rho^+}    \es
\Pi_3^{(1)}(u,s,b)~, \nnb \\
%68
\Pi^{\Xi_b^0 \rar \Lambda_b       \bar{K}^{\ast0}} \es
\Pi_3^{(1)}(u,u,b)~, \nnb \\
%69
\Pi^{\Xi_b^0 \rar \Lambda_b            K^{\ast0}} \es
\Pi_3^{(1)}(u,u,b)~, \nnb \\
%70
\Pi^{\Xi_b^- \rar \Lambda_b            K^{\ast-}} \es
- \Pi_3^{(1)}(u,d,b)~, \nnb \\
%71
\Pi^{\Xi_b^0 \rar \Xi_b^0              \omega}    \es
{1\over \sqrt{2}} \Pi_3^{(1)}(u,s,b)~, \nnb \\
%72
\Pi^{\Xi_b^- \rar \Xi_b^-              \omega} \es
{1\over \sqrt{2}} \Pi_3^{(1)}(d,s,b)~, \nnb \\
%73
\Pi^{\Lambda_b \rar \Lambda_b          \omega}    \es
{1\over \sqrt{2}} \Big[\Pi_3^{(1)}(d,u,b) + \Pi_3^{(1)}(u,d,b) \Big]~, \nnb \\
%74
\Pi^{\Xi_b^0 \rar \Xi_b^0              \phi}    \es
\Pi_3^{(1)}(s,u,b)~, \nnb \\
%75
\Pi^{\Xi_b^- \rar \Xi_b^-              \phi}    \es
\Pi_3^{(1)}(s,d,b)~.
\eaeeq

The expressions for the charmed baryons can easily be obtained by making the
replacement $b \rar c$ and adding to charge of each baryon a positive unit
charge. 

\eAPP

\newpage

\begin{figure}[t]
\begin{center}
\scalebox{0.8}{\includegraphics{fig1.epsi}}
\end{center}
\caption{The dependence of the $f_1^{\Xi_b^{'0} \rar
\Xi_b^{'0} \rho^0}$ on $M^2$ at fixed values of the $s_0$ and $\beta$.}
\end{figure}

\begin{figure}[b]
\begin{center}
\scalebox{0.8}{\includegraphics{fig2.epsi}}
\end{center}
\caption{The dependence of the $f_2^{\Xi_b^{'0} \rar
\Xi_b^{'0} \rho^0}$ on $M^2$ at fixed values of the $s_0$ and $\beta$.}
\end{figure}

\begin{figure}[t]
\begin{center}
\scalebox{0.8}{\includegraphics{fig3.epsi}}
\end{center}
\caption{The dependence of the $f_1^{\Xi_b^{'0} \rar
\Xi_b^{'0} \rho^0}$ on $cos\theta$ at fixed values of the $s_0$ and $M^2$.}
\end{figure}

\begin{figure}[t]
\begin{center}
\scalebox{0.8}{\includegraphics{fig4.epsi}}
\end{center}
\caption{The dependence of the $f_2^{\Xi_b^{'0} \rar
\Xi_b^{'0} \rho^0}$ on $cos\theta$ at fixed values of the $s_0$ and $M^2$.}
\end{figure}


\newpage

\begin{thebibliography}{99}

\bibitem{Rhv01} P. Biassoni,
  arXiv: 1009.2627 (2010).

\bibitem{Rhv02} G. Kane, (ed.), and A. Pierce, (ed.),
  ``Perspectives on LHC physics", (Michigan U.). 2008. 337pp.
    Hackensack, USA: World Scientific (2008) 337 p.

\bibitem{Rhv03} M. A. Shifman, A. I. Vainshtein and V. I. Zakharov,
  Nucl. Phys. B {\bf 147}, 385 (1979).

\bibitem{Rhv04} V. M. Braun,
  prep: hep--ph/9801222 (1998).

\bibitem{Rhv05} T. M. Aliev, K. Azizi, and M. Savc{\i}, Phys. Rev. D {\bf 80}, 096003 (2009).

\bibitem{Rhv06} T. M. Aliev, A. \"{O}zpineci, S. B. Yakovlev, V. Zamiralov,
  Phys. Rev. D {\bf 74}, 116001 (2006).

\bibitem{Rhv07} T. M. Aliev, A. \"{O}zpineci, M. Savc{\i} and V. Zamiralov,
  Phys. Rev. D {\bf 80}, 016010 (2009).

\bibitem{Rhv08} T. M. Aliev, K. Azizi, A. \"{O}zpineci and M. Savc{\i},
  Phys. Rev. D {\bf 80}, 096003 (2009).

\bibitem{Rhv09} T. M. Aliev, A. \"{O}zpineci, M. Savc{\i} and V. Zamiralov,
  Phys. Rev. D {\bf 81}, 056004 (2010).

\bibitem{Rhv10} T. M. Aliev, K. Azizi, and M. Savc{\i},
  Nucl., Phys A, 847 (2010) 101.

\bibitem{Rhv11} P. Ball, V. M. Braun, Y. Koike, and K. Tanaka,
  Nucl. Phys. B {\bf 529}, 323 (1998).

\bibitem{Rhv12} P. Ball, V. M. Braun,
  Nucl. Phys. B {\bf 543}, 201 (1999);
                P. BAll, V. M. Braun, and A. Lenz,                         
  JHEP B {\bf 90}, 0708 (2007).
\bibitem{Rhv13} I. I. Balitsky and V. M. Braun,
  Nucl. Phys. B {\bf 311}, 239 (1988).  

\bibitem{Rhv14} I. I. Balitsky, V. M. Braun, and A. 
                V. Kolesnichenko,
  Nucl. Phys. B {\bf 312}, 509 (1989);
                K. G. Chetyrkin, A. Khodjamirian, and 
                A. A. Pivovarov,
  Phys. Lett. B {\bf 651}, 250 (2008).
                
\bibitem{Rhv15} V. M. Belyaev, V. M. Braun, A. Khodjamirian and 
                R. R\"{u}ckl,
  Phys. Rev. D {\bf 51}, 6177 (1995).

\bibitem{Rhv16} T. M. Aliev, K. Azizi, A. \"{O}zpineci,
  Phys. Rev. D {\bf 79}, 056005 (2009).

\bibitem{Rhv17} P. Z. Huang, H. X. Chen, S. L. Zhu,
  Phys. Rev. D {\bf 80}, 094007 (2009).


\end{thebibliography}
\end{document}